\documentclass[showpacs,preprintnumbers,amsmath,amssymb,superscriptaddress,aip]{revtex4-1}
\usepackage{graphicx}

\def\beq{\begin{equation}}
\def\eeq{\end{equation}}
\def\beqar{\begin{eqnarray}}
\def\eeqar{\end{eqnarray}}

\def\para{\parallel}

\newcommand{\pdiff}[2]{\frac{\partial#1}{\partial#2}}

\newcommand{\pdt}{\partial_t}
\newcommand{\pdr}{\partial_r}

\def\grad{\nabla}

\newcommand{\gradpar}{\grad_\parallel}
\newcommand{\gradperp}{\grad_\perp}

\newcommand{\vpe}{v_{\parallel e}}

\newcommand{\nue}{\nu_{e}}

\newcommand{\nuin}{\nu_{in}}
\newcommand{\kpe}{\kappa_{\parallel e}}

\newcommand{\fmie}{\frac{m_i}{m_e}}

\begin{document}

\title{Nonlinear instability in simulations of Large Plasma Device turbulence}

\author{B. Friedman}
\email{friedman@physics.ucla.edu}

\author{T.A. Carter}

\affiliation{Department of Physics and Astronomy, University of California, Los Angeles, California 90095-1547, USA}

\author{M.V. Umansky}
\affiliation{Lawrence Livermore National Laboratory, Livermore, California 94550, USA}

\author{D. Schaffner}

\affiliation{Department of Physics and Astronomy, University of California, Los Angeles, California 90095-1547, USA}

\author{I. Joseph}

\affiliation{Lawrence Livermore National Laboratory, Livermore, California 94550, USA}

\begin{abstract}
Several simulations of turbulence in the Large Plasma Device (LAPD) [W. Gekelman \emph{et al.}, Rev. Sci. Inst. {\bf 62}, 2875 (1991)] are energetically analyzed and compared with each other
and with the experiment.
The simulations use the same model, but different axial boundary conditions. They employ either periodic, zero-value, zero-derivative, or sheath axial boundaries. The linear stability physics is
different between the scenarios because the various boundary conditions allow the drift wave instability to access different axial structures, and the sheath boundary simulation
contains a conducting wall mode instability which is just as unstable as the drift waves. Nevertheless, the turbulence in all the simulations is relatively similar because it is primarily
driven by a robust nonlinear instability that is the same for all cases.
The nonlinear instability preferentially drives $k_\parallel = 0$ potential energy fluctuations, which
then three-wave couple to $k_\parallel \ne 0$ potential energy fluctuations in order to access the adiabatic response to transfer their energy to kinetic energy fluctuations.
The turbulence self-organizes to drive this nonlinear instability, which destroys the linear eigenmode structures, making the linear instabilities ineffective. 
\end{abstract}

\maketitle

\section{Introduction}

Hydrodynamic turbulence often occurs in the absence of linear instability, e.g. turbulence in pipe flow (Pouseille flow)~\cite{manneville2008}.  Although many robust linear instabilities exist in magnetized plasmas, nonlinear instability can arise as has been shown in many turbulence simulations.
Sometimes the instabilities were found to be of the subcritical~\cite{waltz1985,scott1990,scott1992,nordman1993,itoh1996,highcock2012} or supercritical~\cite{dimits2000,ernst2004} variety, 
while at other times the nonlinear instability simply overpowered a particular linear instability to drive the 
turbulence~\cite{biskamp1995,drake1995,zeiler1996,zeiler1997,korsholm1999,scott2002,scott2003,scott2005,friedman2012b}. In the scenario applicable to turbulence in
the Large Plasma Device (LAPD)~\cite{Gekelman1991} -- in which the magnetic fields lines are straight and without shear -- several studies showed the turbulence to be driven by a nonlinear instability
of this last type, where the nonlinear instability imposed itself over a linear drift-wave instability~\cite{biskamp1995,drake1995,korsholm1999,friedman2012b}. 
One of these~\cite{friedman2012b} used LAPD experimental parameters and profiles in the simulation and demonstrated
that the nonlinear instability was necessary to drive turbulence with characteristics similar to that of the experiment.
Now, these studies~\cite{biskamp1995,drake1995,korsholm1999,friedman2012b} all ascertained
that the mechanism driving the nonlinear instability relies upon axial wavenumber transfers between $k_\para = 0$ and $k_\para \ne 0$ structures. The reason is that the turbulence self-organizes
to preferentially drive $k_\para = 0$ density and temperature fluctuations, taking energy from the density and temperature equilibrium gradients. But in order to access the adiabatic response,
which transfers energy into the dynamically critical $E \times B$ flows, the $k_\para = 0$ fluctuations must transfer their energy through nonlinear three-wave decay into $k_\para \ne 0$ fluctuations.

All of these straight magnetic field simulations, however, employed periodic boundary conditions in the axial (field-aligned) direction.   
One can justifiably question the use of periodic axial boundary conditions to model LAPD and wonder whether the nonlinear instability is an artifact of this choice.  
The use of this boundary condition, for example, prevents the fastest growing linear drift-waves in LAPD from being captured in the simulation: 
with periodic boundary conditions the longest wavelength mode has its wavelength equal to the length of the device whereas the fundamental mode 
(half-wavelength equal to the device length) has a higher growth rate.  This may be the reason why the simulations produce $k_\para = 0$ structures. Would these disappear if the simulations
are allowed access to longer wavelength structures through different boundary conditions?
Furthermore, the actual LAPD axial boundary includes conducting structures; 
it is well known that sheaths on metal walls can drive linear instabilities like the conducting wall mode~\cite{berk1991}. Clearly, periodic boundary simulations miss this wall physics.
Axial boundary conditions do in fact have a significant
affect on the linear stability properties of the system, which is shown in detail in Section~\ref{sec_linear}. 
And since the nonlinear instability relies upon very-long-parallel-wavelength modes to operate, it is 
reasonable to speculate that the boundaries also might have a significant affect on the nonlinear instability properties.

The real axial boundaries in LAPD are complicated. One side of the device contains a hot, emissive cathode behind a mesh anode, and in front of that sits a biased limiter with radius slightly
less than the cathode radius~\cite{schaffner2012}.
The other side contains a floating mesh plate, which is likely shielded by a layer of neutral gas (the plasma may be detached from the plate on this end). Not only is it difficult to determine what the actual axial boundary conditions are, it is also
difficult to develop and implement models for the boundaries. Thus, this paper takes a simpler approach of exploring the affect of 
non-periodic axial boundary conditions on the nonlinear turbulent dynamics using various
idealized boundaries, leaving the calculation and implementation of physically realistic boundary conditions to future work. 
The different axial boundary conditions used here include zero-value, zero-derivative, and perfectly conducting metal plates, which can be modeled with a Bohm sheath condition. 
The main finding is that the nonlinear instability is robust to changes in boundary conditions:  while the linear stability properties are modified significantly with different boundary conditions,  the 
nonlinear instability still dominates the turbulent drive in all cases. 
In fact, the qualitative properties of the turbulence and the turbulent
dynamics are similar between the simulations. Quantitatively, there are some differences between the simulations such as varying fluctuation levels and varying degrees to which
the nonlinear instability dominates the linear ones.

The paper is organized as follows: Section~\ref{dw_model} presents the model and boundary conditions used in the simulations, while 
Sec.~\ref{sec_linear} goes over the origins and properties of the linear instabilities in the different simulations.
Section~\ref{sec_energetics_machinery} develops the energetics equations that are used in Sec.~\ref{dyn_results} to reveal the details of the nonlinear instability in the simulations.
Finally, Sec.~\ref{Sec_lin_vs_nl} explores the effect of the nonlinear instability on the mode structure of the turbulence.

\section{The Simulation Model}
\label{dw_model}

A Braginskii-based fluid model~\cite{Braginskii1965} is used to simulate global drift wave turbulence in LAPD using the BOUT++ code~\cite{dudson2009}. 
The evolved variables in the model are the plasma density, $N$, the electron fluid parallel velocity $\vpe$, the potential vorticity $\varpi \equiv \gradperp \cdot (N_0 \gradperp \phi)$,
and the electron temperature $T_e$. The ions are assumed cold in the
model ($T_i = 0$), and sound wave effects are neglected ($v_i = 0$). Details of the simulation code, derivations of the model, grid convergence studies, and analyses of simplified models
may be found in previously published LAPD simulation studies~\cite{Popovich2010a,Popovich2010b,Umansky2011,friedman2012,friedman2012b}.

The equations are developed with Bohm normalizations: lengths are
normalized to the ion sound gyroradius, times to the ion
cyclotron time, velocities to the sound speed, densities to the equilibrium peak density, and electron
temperatures and potentials to the equilibrium peak electron temperature. These normalizations are constants (not functions of radius) and are calculated from these reference values:
the magnetic field is $1$ kG, the ion unit mass is $4$, the peak density is $2.86 \times 10^{12}$ cm$^{-3}$, and the peak electron temperature
is $6$ eV. The equations are:

\beqar
\label{ni_eq}
\pdt N = - {\mathbf v_E} \cdot \grad N_0 - N_0 \gradpar \vpe + \mu_N \gradperp^2 N + S_N + \{\phi,N\}, \\
\label{ve_eq}
\pdt \vpe = - \fmie \frac{T_{e0}}{N_0} \gradpar N - 1.71 \fmie \gradpar T_e  + \fmie \gradpar \phi - \nue \vpe + \{\phi,\vpe \}, \\
\label{rho_eq}
\pdt \varpi = - N_0 \gradpar \vpe  - \nuin \varpi + \mu_\phi \gradperp^2 \varpi + \{\phi,\varpi \}, \\
\label{te_eq}
\pdt T_e = - {\mathbf v_E} \cdot \grad T_{e0} - 1.71 \frac{2}{3} T_{e0} \gradpar \vpe + \frac{2}{3 N_0} \kpe \gradpar^2 T_e  \nonumber \\
- \frac{2 m_e}{m_i} \nue T_e  + \mu_T \gradperp^2 T_e +  S_T + \{\phi,T_e\}.
\eeqar

\begin{figure}[!htbp]
\includegraphics[]{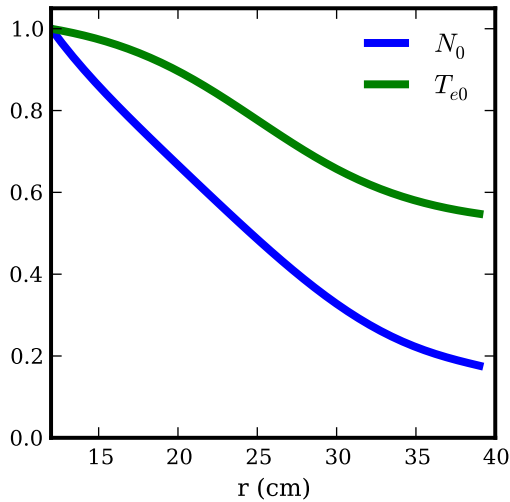}
\hfil
\caption{The profiles of density $N_0$ and electron temperature $T_{e0}$ used in the simulations normalized to their peak values of $2.86 \times 10^{12}$ cm$^{-3}$ and 
$6$ eV, respectively.}
\label{eq_profiles}
\end{figure}

In these equations, $\mu_N$, $\mu_T$, and $\mu_\phi$ are artificial diffusion and viscosity coefficients used for subgrid dissipation. They are large enough to allow saturation
and grid convergence~\cite{friedman2012}, but small enough to allow for turbulence to develop. In the simulations, they are all given the same value of $1.25 \times 10^{-3}$ in Bohm-normalized units. 
This is the only free parameter in the simulations. All other parameters such as the electron collisionality $\nue$, ion-neutral
collisionality $\nuin$, parallel electron thermal conductivity $\kpe$, and mass ratio $\fmie$ are calculated from experimental quantities.
There are two sources of free energy: the density gradient due to the equilibrium density profile $N_0$, and the equilibrium electron temperature gradient in $T_{e0}$, both of which are
taken from experimental fits. $N_0$ and $T_{e0}$ are functions of only the radial cylindrical coordinate $r$, and they are shown in Fig.~\ref{eq_profiles}. 
The mean potential profile $\phi_0$ is set to zero in the model, and terms involving $\phi_0$ are not included in Eqs.~\ref{ni_eq}-\ref{te_eq}. 
The justification for this is that biasable azimuthal limiters in LAPD allow for the mean $E \times B$ flow and flow shear to be varied with high precision, even allowing the flow to be
nulled out~\cite{schaffner2012}. 
The simulations in this paper use the $N_0$, $T_{e0}$, and $\phi_0$ profiles from the nulled out flow experiment, justifying setting the mean potential profile to zero in the simulations.

Simulations also use density and temperature sources ($S_n$ and $S_T$) in order to keep the equilibrium profiles from relaxing away from their experimental shapes. 
These sources suppress the azimuthal averages ($m=0$ component of the density and temperature fluctuations) at each time step. 
The azimuthal average of the potential $\phi$ is allowed to evolve in
the simulation, allowing zonal flows to form, although they are relatively unimportant to the turbulent dynamics~\cite{friedman2012b}.

The terms in Poisson brackets are the $E \times B$ advective nonlinearities, which are the only nonlinearities used in the simulations.
The numerical simulations use finite differences in all three dimensions and use cylindrical annular geometry ($12<r<40$ cm). The radial extent used in the simulation
encompasses the region where fluctuations are above a few percent in the experiment. Therefore, the radial boundaries are fixed to zero value. The grid contains 128 radial points,
1024 azimuthal points, and 32 axial points. The grid resolves $\rho_s$ in the perpendicular plane and allows for spatial convergence of the solution~\cite{friedman2012}.

This study analyzes five turbulent simulations which will be referred to as (1) the periodic simulation, (2) the $n=0$ suppressed simulation ($n$ is the axial wavenumber), 
(3) the sheath simulation, (4) the Dirichlet simulation,
and (5) the Neumann simulation.
The periodic and $n=0$ suppressed simulations were also analyzed in a previous paper~\cite{friedman2012b}. Both of these simulations enforce periodic boundary conditions in the axial ($z$)
direction. The $n=0$ suppressed simulation adds an artificial sink-like contribution to Eqs.~\ref{ni_eq}-\ref{te_eq} which removes the axial average 
($k_\parallel = 0$ contribution, where $n = k_\para l_\para /2 \pi$) of the fluctuations at each time step. This $n=0$ suppression eliminates nonlinear instability drive and allows
the linear instability to take over the turbulent drive~\cite{friedman2012b}. The $n=0$ suppressed simulation, therefore, serves as a contrast to the periodic simulation in which a nonlinear instability
drives the turbulence.
  
The sheath simulation, as its name implies, uses sheath boundary conditions on the axial machine ends. Specifically, the sheath boundary
condition for the parallel current is a linearized Bohm condition:

\beq
\label{sheath_bndry}
J_\para = N_0 (\phi + \text{log} \sqrt{4 \pi m_e/m_i} \ T_e) 
\eeq

where $J_\para = - N_0 \vpe$. The axial boundary for $\phi$ is set using this relation along with Ohm's law: $ \fmie \gradpar \phi = \nue \vpe$. 
The axial boundaries for the density and temperature fields are implemented with zero-second-derivative boundary conditions. This is somewhat arbitrary, and it is noted
that stringent analytical and numerical calculations have recently been made for such fields in the magnetic pre-sheath region~\cite{loizu2012}, but those have not been applied in this simulation.
The fourth (Dirichlet) simulation uses fixed zero-value axial boundary conditions, while the fifth (Neumann) employes zero-first-derivative conditions to all fields.

\begin{figure}[!htbp]
\includegraphics[]{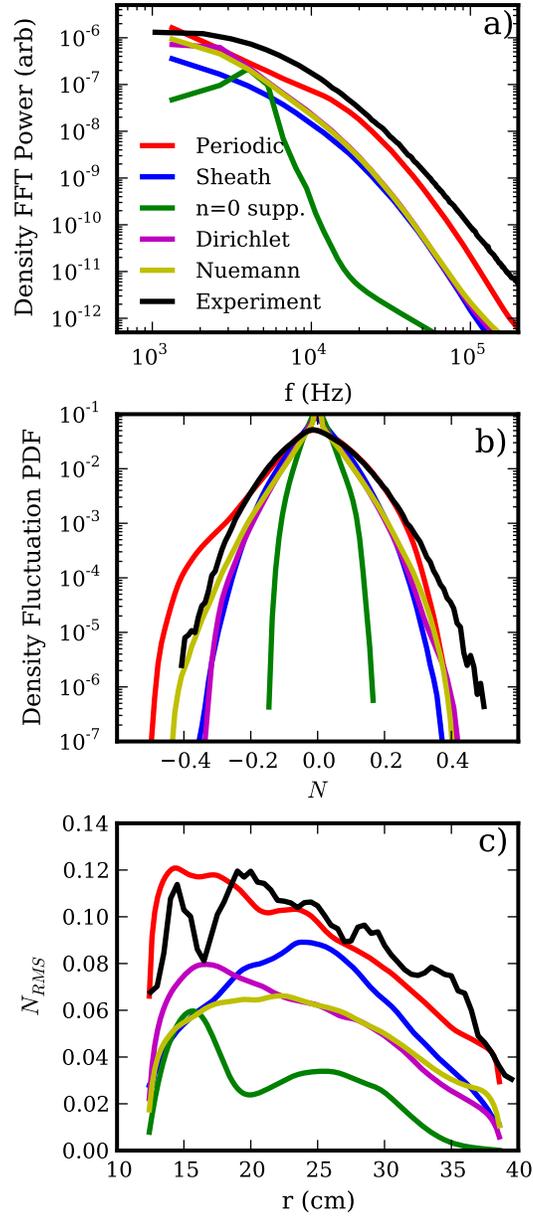}
\hfil
\caption{A comparison of the statistical turbulent properties of the density field. The figures show \textbf{a)} the frequency spectrum, \textbf{b)} the pdf and \textbf{c)} the RMS level
of the fluctuations as a function of radius. The data are calculated from the experiment (black), periodic simulation (red), 
the sheath simulation (blue), the $n=0$ suppressed simulation (green), the Dirichlet simulation (magenta), and the Neumann simulation (yellow). This color scheme is consistent throughout the paper.}
\label{statistics}
\end{figure}

As a first comparison, Fig.~\ref{statistics} shows a few statistical characteristics of the density fluctuations for each of the five simulations along with the corresponding
characteristics from the experiment. The periodic simulation clearly has the most similar characteristics to those of the experiment while the $n=0$ suppressed simulation is most dissimilar.
The fluctuations of the sheath, Dirichlet, and Neumann simulations have similar statistical properties as those from the experiment, however, 
the amplitude of the fluctuations is a bit smaller than the experimental fluctuations in general.

\section{Linear Instabilities}
\label{sec_linear}

\begin{figure}[!htbp]
\includegraphics[width=0.6\textwidth]{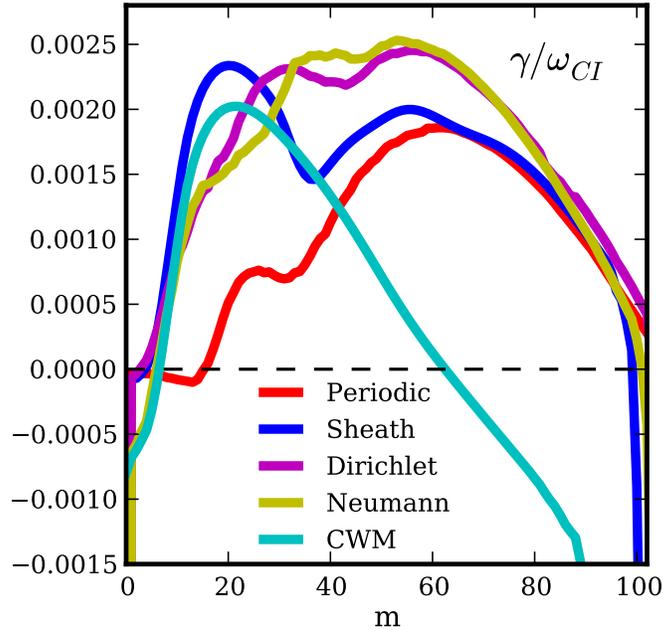}
\hfil
\caption{The linear growth rates of the four simulations (the $n=0$ suppressed simulation has the same linear growth rate as the periodic simulation) along with the growth rate of 
the conducting wall mode (CWM) as a function of azimuthal wavenumber $m$.}
\label{drift_cwm_gamma}
\end{figure}

The model described by the equations of Section~\ref{dw_model} contains a few linear instabilities that can all act at the simulated scales. Two of these instabilities are of the electrostatic
drift wave type - one is driven by the density gradient and the other by the electron temperature gradient. Both of these instabilities supply energy to the electrostatic potential through parallel
compression, called the adiabatic response, and are made unstable by the electron-ion collisional dissipation. These instabilities act under all choices of parallel boundary conditions.

The other instability is called the conducting wall mode (CWM) since it is driven by the conducting wall sheaths on the parallel boundaries. Various terms in Eqs.~\ref{ni_eq}-\ref{te_eq} 
can be eliminated to isolate the conducting wall mode. The reduced set of linearized equations is:

\beqar
\label{ve_eq2}
\pdt \vpe = \fmie \gradpar \phi - \nue \vpe, \\
\label{rho_eq2}
\pdt \varpi = - N_0 \gradpar \vpe - \nuin \varpi + \mu_\phi \gradperp^2 \varpi, \\
\label{te_eq2}
\pdt T_e = - {\mathbf v_E} \cdot \grad T_{e0} + \frac{2}{3 N_0} \kpe \gradpar^2 T_e  - \frac{2 m_e}{m_i} \nue T_e  + \mu_T \gradperp^2 T_e,
\eeqar

along with the axial boundary condition given in Eq.~\ref{sheath_bndry}. The free energy source for this instability is the electron temperature gradient, which is also the free energy
for the thermally driven drift waves. However, the adiabatic response is replaced here with a coupling to the potential through the axial boundary condition.

For the experimental parameters and profiles used in the turbulent simulations, 
the linear growth rates for the drift waves and the CWM are comparable. The growth rates of the fastest growing linear eigenmode at a given azimuthal wavenumber
are shown in Fig.~\ref{drift_cwm_gamma}.
The periodic, sheath, Dirichlet, and Neumann growth rate curves are found by simulating the linearized versions of Eqs.~\ref{ni_eq}-\ref{te_eq} with their respective
axial boundary conditions in BOUT++. Therefore, both the density- and temperature-driven drift wave contributions are present for these curves.
The CWM curve is obtained by simulating Eqs.~\ref{ve_eq2}-\ref{te_eq2} with the sheath axial boundary condition of Eq.~\ref{sheath_bndry}, so there is no drift wave contribution to this curve
due to the absence of the adiabatic response in these equations. 

The linear growth rates for the Dirichlet and Neumann simulations are markedly different from those of the periodic simulation because the zero-value and zero-derivative boundary conditions
allow for more freedom of the axial structure. In other words, they allow for the axial wavenumber $n \equiv k_\para l_\para / 2 \pi$ to take on non-integer values (about $1/2$ in this case) that
are more unstable than the $n=1$ eigenmodes that are enforced by the periodic boundaries. The sheath boundary condition has this affect as well, but more importantly
it affects the linear stability properties of the system at low azimuthal wavenumber $m$ due to the presence of the CWM contribution. The CWM is more unstable
than the drift waves for $m \lessapprox 30$ but less unstable for $m \gtrapprox 30$, and the eigenmode with the highest growth rate has $m=20$ for the sheath simulation as opposed to $m=60$ for the pure
drift wave simulations.
The linear sheath eigenmodes also reflect which of the linear instabilities is active at which wavenumber. In other words, the sheath eigenmodes have CWM character at $m \lessapprox 30$ and drift wave
character at $m \gtrapprox 40$. This manifests itself as differences in axial and radial structure as well as in phase relations between the different scalar fields. However, the linear differences
are only significant in the end if they affect the turbulence dynamics. And following a previous paper~\cite{friedman2012b}, an effective way to study the turbulence dynamics is with
an energy dynamics analysis of the turbulent simulations.

\section{Energetics Formalism}
\label{sec_energetics_machinery}

In order to perform an energy dynamics analysis on the simulations, expressions for the energy and energy evolution must be derived from Eqs.~\ref{ni_eq}-\ref{te_eq}.
To start, an expression for the normalized energy of the wave fluctuations in the model is defined as:

\beq
\label{energy_eq}
E = \frac{1}{2} \int_V  \left[ P_0 \left((N/N_0)^2 + \frac{3}{2} (T_e/T_{e0})^2 \right) + N_0 \left( \frac{m_e}{m_i} \vpe^2 + (\gradperp \phi)^2 \right) \right] dV,
\eeq

where $P_0 = N_0 T_{e0}$ is the equilibrium pressure.
The $\frac{1}{2} P_0 (N/N_0)^2$ term is the potential energy due to density fluctuations, $\frac{3}{4} P_0 (T_e/T_{e0})^2$ is the electron temperature fluctuation potential energy,
$\frac{1}{2} N_0 \frac{m_e}{m_i} \vpe^2$ is the parallel electron kinetic energy, and $\frac{1}{2} N_0 (\gradperp \phi)^2$ is the $E \times B$ perpendicular kinetic energy.

A more detailed look at the energetic processes comes from a spectral energy analysis. To do this, each fluid field $(N,T_e,\vpe,\phi)$ at a given time is Fourier decomposed as 
$F(r,\theta,z) = \sum_{\vec{k}} f_{\vec{k}}(r) e^{i (m \theta + k_z z )}$,
where the subscript $\vec{k}$ represents the spectral wavenumbers, $(m,n)$, and both positive and negative wavenumbers are included in the sums. 
$m$ is the azimuthal wavenumber while $n$ is the axial integer wavenumber. 
Note that the radial direction is not spectrally decomposed because it's not essential here.
With this, the energy of each Fourier $\vec{k} = (m,n)$ mode is

\beq
\label{E_k}
E_{tot}(\vec{k}) = \frac{1}{2} \left< \frac{T_{e0}}{N_0} |n_{\vec{k}}|^2 + \frac{3 N_0}{2 T_{e0}} |t_{\vec{k}}|^2 + \frac{m_e}{m_i} N_0 |v_{\vec{k}}|^2 + N_0 \left| \pdiff{\phi_{\vec{k}}}{r} \right|^2 + N_0 \frac{m^2}{r^2} |\phi_{\vec{k}}|^2 \right>,
\eeq

where the brackets $\left< \right>$ represent the radial integral: $\int_{r_a}^{r_b} r dr$. 
The energy evolution for each Fourier mode of each field has the form:

\beq
\label{dEdt_j}
\pdiff{E_{j}(\vec{k})}{t} = Q_{j}(\vec{k}) + C_{j}(\vec{k}) + D_j(\vec{k}) + \sum_{\vec{k}'} T_{j}(\vec{k},\vec{k}').
\eeq

The index $j$ stands for each field, ($N,T,v,\phi$), and the sum over $j$ gives the total energy evolution. 
The derivation of Eq.~\ref{dEdt_j} is given in the previous work~\cite{friedman2012b}. $T_{j}(\vec{k},\vec{k}')$ is the nonlinear energy transfer function that comes from the advective
nonlinearities.  It describes the nonlinear energy transfer rate of fluctuations with $\vec{k}'=(m',n')$ to fluctuations with $\vec{k}=(m,n)$. 
For example, a positive value of $T_{N}(\vec{k},\vec{k}')$ indicates that density fluctuations
at wavenumber $\vec{k}$ gain energy from density fluctuations at wavenumber $\vec{k}'$, with the process mediated by flow fluctuations at wavenumber $\vec{k}-\vec{k}'$.

The linear terms are broken up into three contributions in Eq.~\ref{dEdt_j}.
$D_{j}(\vec{k})$ represents energy dissipation due to collisions, artificial diffusion and viscosity, and the density and temperature sources.
Each contribution to $D_j(\vec{k})$ is negative. 
$C_j(\vec{k})$ contains the linear terms dubbed ``transfer channels''~\cite{scott2002}. They are:

\beqar
C_N(\vec{k}) & = & Re \left\{ \left< - i k_z T_{e0} v_{\vec{k}} n_{\vec{k}}^* \right> \right\}
\label{Cnk} \\
C_v(\vec{k}) & = & Re \left\{ \left< - i k_z T_{e0} n_{\vec{k}} v_{\vec{k}}^* + i k_z N_0 \phi_{\vec{k}} v_{\vec{k}}^*  - 1.71 i k_z N_0 t_{\vec{k}} v_{\vec{k}}^*  \right> \right\}
\label{Cvk} \\
C_\phi(\vec{k}) & = & Re \left\{ \left< i k_z N_0 v_{\vec{k}} \phi_{\vec{k}}^* \right> \right\}
\label{Cpk} \\
C_T(\vec{k}) & = & Re \left\{ \left< - 1.71 i k_z N_0 v_{\vec{k}} t_{\vec{k}}^* \right> \right\}
\label{Ctk}
\eeqar

First, note that the real part operators are written explicitly in these expressions since the imaginary part of these expressions would cancel with the imaginary part of the 
corresponding terms with $-\vec{k}$. Second, notice that $C_N(\vec{k}) + C_v(\vec{k}) + C_\phi(\vec{k}) + C_T(\vec{k}) = 0$.
This is the reason why these terms are called transfer channels. They represent the transfer
between the different types of energy of the different fields ($N,\phi,T_e \leftrightarrow v_{\para e}$), but taken together, they do not create or dissipate total
energy from the system. The only energy transfer between different fields in this system is through the parallel electron velocity (parallel current) dynamics. There is no direct transfer between
the state variables $N, \phi,$ and $T_e$.  Altogether, the coupling through the parallel current is called the
adiabatic response. It is an essential part of both the linear and nonlinear
drift wave mechanisms~\cite{scott2002,scott2005}. The adiabatic response moves energy from the pressure fluctuations to the perpendicular flow through the parallel current.

Finally, the $Q_j(\vec{k})$ terms represent the fluctuation energy sources. They are:

\beqar
Q_N(\vec{k}) & = & Re \left\{ \left< -\frac{i m T_{e0}}{N_0 r} \pdr N_0 \phi_{\vec{k}} n_{\vec{k}}^*  \right> \right\}
\label{Qnk} \\
Q_v(\vec{k}) & = & 0
\label{Qvk} \\
Q_\phi(\vec{k}) & = & 0
\label{Qpk} \\
Q_T(\vec{k}) & = & Re \left\{ \left< -\frac{3}{2} \frac{i m N_0}{T_{e0} r} \pdr T_{e0} \phi_{\vec{k}} t_{\vec{k}}^*  \right> \right\}
\label{Qtk}
\eeqar

$Q_N(\vec{k})$ is the energy extraction from the equilibrium density profile into the density fluctuations. 
This term may have either sign depending on the phase relation between $\phi_{\vec{k}}$ and $n_{\vec{k}}$, 
so it can in fact dissipate fluctuation potential energy from the system as well as create it
at each $\vec{k}$. $Q_T(\vec{k})$ is completely analogous to $Q_N(\vec{k})$ but for the temperature rather than the density. 
$Q_v(\vec{k})$ and $Q_\phi(\vec{k})$ are zero because the parallel and perpendicular flow fluctuations obtain energy only through the adiabatic response, not directly through the free
energy in the equilibrium gradients.

\section{Energy Dynamics Results}
\label{dyn_results}

\begin{figure}[!htbp]
\includegraphics[width=0.85\textwidth]{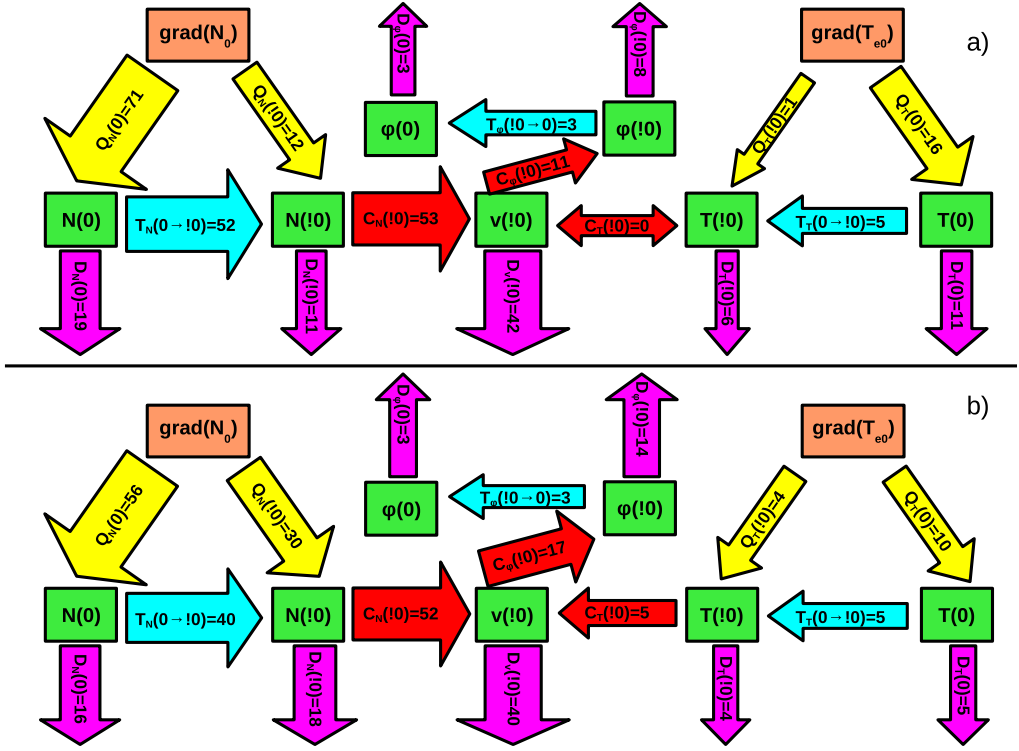}
\hfil
\caption{Summary of the energy dynamics for the \textbf{a)} periodic case and \textbf{b)} sheath boundary case. Each arrow contains the sum over all $m$. The 0's in the parentheses
represent the $n=0$ modes while the !0's represent a sum over the $n$ modes for $n \ne 0$. The values in the arrows represent the percentage of total energy that goes through the channel
represented by the arrow. The size of the arrows gives a rougher but more visual indication of the amount of energy going through each channel.}
\label{en_diagrams}
\end{figure}

The diagrams in Fig.~\ref{en_diagrams} summarize the flow of energy for the periodic and sheath simulations. Each of the functions, such as $Q_N(m,n)$, is a function of $m$ and $n$, making
visualization of all of these functions difficult. So the terms in the diagrams are summed over $m$. Additionally, all of the $n \ne 0$ terms are summed over as well.
The $n=0$ contribution is separated from the other $n$ components because
the $n=0 \leftrightarrow n \ne 0$ dynamic is the primary factor that determines whether the linear instability or the nonlinear instability dominates the energy drive~\cite{friedman2012b}.

In these diagrams, the source of energy into the fluctuations is free energy in the equilibrium gradients, 
$\nabla N_0$ and $\nabla T_{e0}$. The arrows labeled $Q_N$ and $Q_T$ represent energy injection from
the equilibrium gradients into the fluctuations, $n(\vec{k})$ and $t(\vec{k})$. The four $Q$ arrows contain values that sum to $100$ (by choice of normalization). 
Since the $Q$ pathways are the only pathways that deposit
net energy into the fluctuations, the numbers in all arrows represent a percentage of the total energy injected into the system.
Now, a majority of the energy deposited into the fluctuations ($71\%$ for the periodic simulation and $56\%$ for the sheath simulation)
is from the density gradient into the $n=0$ density fluctuations. This is not a path allowed by the linear drift-wave instabilities in the system since
they can only deposit energy into $n \ne 0$ fluctuations.
In fact, in this turbulent state, more energy is transfered by nonlinear three-wave coupling into $n \ne 0$ fluctuations than by direct injection from the equilibrium gradients. The three-wave
coupling is represented by the $T_N$ and $T_T$ arrows. The direction of these arrows is from $n=0 \rightarrow n \ne 0$, which is opposite to that expected from the common cascading type 
turbulent paradigm where the linear instability dominates the turbulent injection dynamics and three-wave processes transfer energy to waves that are linearly stable.

The reason why all of the non-dissipated energy that is injected into $n=0$ density and temperature structures goes into 
$n \ne 0$ density and temperature potential energy structures rather than into $n=0$
kinetic energy structures is that potential to kinetic energy transfer can only work through the adiabatic response, which requires $n \ne 0$. Actually, in the sheath simulation, potential energy
can transfer to kinetic energy through the axial boundaries, but this still requires $n \ne 0$ and it works only through the temperature fluctuations, which are less important than the density
fluctuations in the simulations. Note, in fact, that boundary contributions aren't included in the energy dynamics calculations as they are insignificant.
So the main transfer channel from potential to kinetic energy, shown by the $C_N$, $C_T$, and $C_\phi$ arrows, is the adiabatic response. The adiabatic response takes energy
from the $n \ne 0$ density and temperature fluctuations and transfers it into the parallel velocity fluctuations ($C_N$ and $C_T$, respectively). It then transfers some of this energy into
$n \ne 0$ potential fluctuations ($C_\phi$), while much of the fluctuation energy is ohmically dissipated ($D_v$).

The final step of the energy dynamics process is a three-wave axial wavenumber transfer from potential fluctuations at $n \ne 0$ to potential fluctuations at $n=0$, which are of course neccessary
in making the $Q_N(0)$ and $Q_T(0)$ terms finite. Meanwhile, dissipation acts on all fluctuations, which is quantified by the $D$ arrows throughout. A couple of interesting features are evident
from the diagrams in Fig.~\ref{en_diagrams}. 
First, the turbulent dynamics in both simulations are dominated by the nonlinear instability process described above and in Friedman et al.~\cite{friedman2012b}
rather than the paradigmatic process of linear instability energy injection followed by nonlinear cascading. And second,
the periodic and sheath simulations have qualitatively similar dynamics despite the fact that the linear stability properties of the two cases
are qualitatively different. This is also true of the Dirichlet and Neumann simulations, although their diagrams are not shown in Fig.~\ref{en_diagrams}. 
This speaks to the robustness of the nonlinear instability.

\begin{figure}[!htbp]
\includegraphics[]{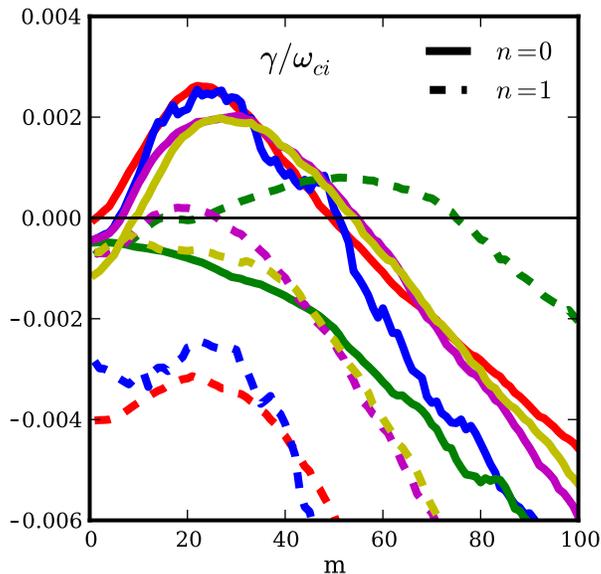}
\hfil
\caption{ The turbulent growth rates with $\gamma$ defined in Eq.~\ref{gamma_def}. The $n=0$ (solid) and $n = 1$ (dashed) 
growth rates are displayed as a function of $m$. The different colors represent different simulations, consistent with the scheme used in Fig.~\ref{statistics}}
\label{nl_vs_lin_gamma}
\end{figure}

A more compact way to see the similarity of the instability process in all of the simulations is shown in Fig.~\ref{nl_vs_lin_gamma}. 
This figure shows the turbulent growth rate of the five simulations. 
The turbulent growth rate is defined as the net energy injection
into the fluctuations minus the dissipation out of them, all divided by the total energy. The conservative transfers ($C$'s and $T$'s) are of course not part of the growth rates.
Formally, 

\beq
\label{gamma_def}
\gamma(\vec{k}) = \left[ \sum_j Q_j(\vec{k}) + D_j(\vec{k}) \right] /2 E_{tot}(\vec{k}), 
\eeq

where the index j represents the different fields.
Since the growth rates sum over the fields, it's not as difficult to view them in their full wavenumber space (both in $m$ and $n$). However, almost all of the energy in the simulations is 
contained in $n=0$ and $n=1$ fluctuations, so $n>1$ fluctuations are not shown in the figure. It should be noted, however, that Fourier decomposing the non-periodic simulations (sheath, Dirichlet, and Neumann)
in the axial direction is less ideal than doing so in the periodic simulations (periodic and $n=0$ suppressed). 
Fourier modes are not as natural a basis in the non-periodic simulations, and the $n=1$ Fourier mode does not perfectly represent the linear
eigenmode structure as it does for the periodic simulations. 
A more detailed discussion of this point is left to the Appendix, but in summary, the $n=1$ Fourier mode does capture enough of the
linear eigenmode structures of the non-periodic simulations to make the Fourier decomposition useful for this simulation.

Figure~\ref{nl_vs_lin_gamma} illustrates the true dominance of the nonlinear instability in the generally positive $n=0$ energy growth rate and generally negative $n=1$ growth rate for all of the simulations
except the $n=0$ suppressed simulation. Take the periodic simulation as the clearest example of this point since there is no ambiguity in the Fourier transform for this simulation. 
Fig.~\ref{drift_cwm_gamma} shows the \emph{linear} growth rate as a function of $m$ only, but implicitly, $n=1$ for the periodic curve. The reason is that the linear eigenmodes of the periodic simulation
are Fourier modes and all eigenmodes with $n>1$ have smaller $\gamma_{lin}(m)$ 
than the $n=1$ eigenmodes and all $n=0$ flute eigenmodes have large negative $\gamma_{lin}(m)$ because the linear instability mechanism doesn't work when $n=0$. On the other hand, the \emph{turbulent}
growth rate curves in Fig.~\ref{nl_vs_lin_gamma} for the periodic simulation have a very different nature than the corresponding linear growth rate curves.
For low $m$, the $n=0$ turbulent growth rate is positive, while the $n=1$ turbulent growth rate is negative for all $m$. This reversal in sign is indicative of the nonlinear instability.
The sheath, Dirichlet, and Neumann simulations are not as easy to analyze because when their most unstable linear structures are axially Fourier decomposed, they contain pieces of all $n$ components
(their eigenmodes are not Fourier modes).
However, it's clear that the turbulent growth rates are quite similar between these simulations and to the periodic one, indicating that the same nonlinear instability mechanism is acting in each of
these cases as well. The similarity in the turbulent structures (see Fig.~\ref{statistics}) also points to this conclusion.
This is in stark contrast to the $n=0$ suppressed simulation, which has an $n=1$
growth rate similar to the linear growth rate (Fig.~\ref{drift_cwm_gamma}) and a negative $n=0$ growth rate. Note that even though the $n=0$ fluctuation components are 
removed from the $n=0$ suppressed simulation at each time
step, $n=0$ fluctuations are nonlinearly excited (by three-wave transfer) by the $n \ne 0$ fluctuations, 
and therefore, they do have small but finite amplitude prior to their removal, which can be used to
calculate the growth rate of these modes. Furthermore, the turbulent growth rate of the $n=1$ component of the $n=0$ suppressed simulation is
slightly less than the linear growth rate due to the fact that eigenmodes other than the fastest growing ones are nonlinearly excited in the turbulent simulation, 
thus damping the growth rate.
But this $n=0$ suppressed growth rate picture is just that of the turbulence paradigm of linear instability with cascading dynamics, which is significantly different than the nonlinear
instability picture of the other simulations.

\section{Linear vs Nonlinear Structure Correlation}
\label{Sec_lin_vs_nl}

Now it may be the case that in simulations dominated by a linear instability, the fastest growing linear
eigenmode dominates the system, nonlinearly transfering some energy to more weakly unstable or even stable eigenmodes. In this case, a large portion of the energy may remain in the fastest
growing linear eigenmode~\cite{hatch2011}. In the case where a nonlinear instability is dominant, the linear eigenmode should have little bearing on the structure of the turbulence and therefore little
energy should be contained in this eigenmode. Therefore, a gauge of whether a linear or nonlinear instability dominates a system is the fraction of energy in a turbulent system
that is contained in the fastest growing linear eigenmode. This may be calculated by projecting the fastest growing eigenmode onto the turbulent state.

Formally, in the model considered in this study, the turbulent state is fully described by four independent fields, which can be appended into a single vector of the spatio-temporal field functions: 
$f_{turb}(\vec{r},t) = \{N(\vec{r},t),T_e(\vec{r},t),\gradperp \phi(\vec{r},t), \vpe(\vec{r},t)\}$. This vector may be decomposed in a complete basis:

\beq
\label{basis_decomp}
f_{turb}(\vec{r},t) = \sum_{i,m} c_{i,m}(t) \psi_{i,m}(r,z) e^{i m \theta},
\eeq

where $\psi_{i,m}(r,z)$ are time-independent spatial complex basis functions of the form $\psi_{i,m}(r,z) = \left\{ n_{i,m}(r,z),t_{i,m}(r,z),\gradperp \phi_{i,m}(r,z), v_{i,m}(r,z) \right\}$,
and $c_{i,m}(t)$ are the complex time-dependent amplitudes. The $\theta$ dependence of the basis functions has been explicitly imposed as a Fourier basis. The total number of
linearly independent basis functions is the number of total grid points used in the simulation times the number of independent fields, which is four in this case.
Now, $\psi_{i,m}(r,z)$ can be any linearly independent set of functions and need not be the linear eigenfunctions
of the system. In fact, the linear eigenfunctions of the equations used here are not orthogonal, and are thus not very useful to consider. 
However, it is quite useful to set $\psi_{0,m}(r,z)$ to the fastest
growing linear eigenmode because this is the structure of interest that is to be projected onto the turbulence. 
The other $\psi_{i \ne 0,m}(r,z)$ comprise the remainder of the orthonormal basis, and they must be different from
the remaining linear eigenfunctions in order to complete the orthogonal basis. It isn't necessary for the purpose of this study to actually compute these other basis functions, but if one were to compute
them, one might start with all of the linear eigenmodes
and perform a Gram-Schmidt orthogonalization procedure, making sure to start with the fastest growing eigenmode in order to preserve it. Using this procedure, Hatch et al.~\cite{hatch2011}
found that a significant fraction ($\sim 50\%$) of the energy in a turbulent state of ITG turbulence was contained in the fastest growing linear eigenmode at each perpendicular wavenumber.
Such a result, however, doesn't require knowledge of the other basis functions, and thus they are not computed here.

Now, to compute this fraction, first define an inner product that is energetically meaningful and that sets the orthonormality of the basis functions:

\beq
\label{basis_orthonormality}
\left< \psi_{i,m},\psi_{j,m} \right> = \int w \psi_{i,m}^* \cdot \psi_{j,m} dV = \delta_{i,j}.
\eeq

The weighting $w$ is such that $\left< f_{turb}, f_{turb} \right> = E_{turb}$.
Now from Eqs.~\ref{basis_decomp} and~\ref{basis_orthonormality}, $\left< f_{turb}, f_{turb} \right> = E_{turb} = \sum_{i,m} |c_{i,m}|^2$ and 
$\left< f_{turb,m}, f_{turb,m} \right> = E_{turb,m} = \sum_i |c_{i,m}|^2$.
Then, the amount of energy contained in the fastest growing mode (for each $m$) is given by the square of the projection
of the mode onto the turbulence: $E_{0,m} = \left| \left< \psi_{0,m}, f_{turb,m} \right> \right|^2 = |c_{0,m}|^2$. The ratio 
$R_m = E_{0,m}/E_{turb,m}$ is a measure of the fraction of turbulent energy contained in the fastest growing linear eigenmode. 

Of course, $E_{turb,m}$ is easily calculated from the turbulent state, but $E_{0,m}$ in the 
turbulent state can only be found with knowledge of the fastest growing eigenfunction. The fastest growing eigenfunction, though, can be found easily by running a simulation from a random 
or turbulent state with all of the nonlinearities removed from the model equations as was done in Section~\ref{sec_linear}. After some time, the fastest growing eigenfunctions will come to
dominate the fluctuation structure. Then, a Fourier decomposition in $m$ space will separate the fastest growing eigenfunctions at each $m$, including the real and imaginary part
of the eigenfunctions (up to a time dependent complex constant, which is removed by normalizing the eigenfunction). These eigenfunctions can then be projected onto the turbulent state
with the inner product defined in Eq.~\ref{basis_orthonormality}, giving $E_{0,m}$.

\begin{figure}[!htbp]
\includegraphics[]{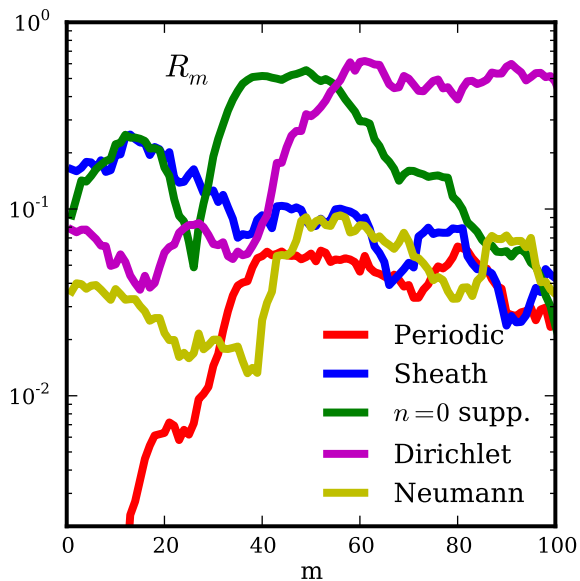}
\hfil
\caption{The ratio $R_m$ of the turbulent energy in the fastest growing linear eigenmode to the total energy at each $m$ for all of the simulations.}
\label{ratios}
\end{figure}

The ratio $R_m$ is shown in Fig.~\ref{ratios} for the five simulations. For the most part, the simulations other than the $n=0$ suppressed one have a small value of the ratio ($R_m < 0.25$) for all $m$. 
This confirms that the turbulence largely self-organizes
without regard to the linear physics. The one exception is the Dirichlet simulation for $m > 50$, which has $R_m \sim 0.5$. Most of the energy in this and the other simulations, however, 
is at low $m$ where the turbulent growth rates are largest~\cite{friedman2012b}, so these larger $m$ eigenmodes don't make a large impact on the overall structure of the turbulence.
In fact, $R_m$ is below $0.1$ for $m<40$ for the periodic, Dirichlet, and Neumann simulations, precisely the area where
$n=0$ structures dominate the energy spectrum~\cite{friedman2012b}. It is not surprising then that the fastest growing linear eigenfunctions, which have little or no flute character,
don't make up much of the energy of the signal. The sheath simulation shows more linear eigenmode dominance at low $m$ because of the relatively large CWM growth rate, but $R_m$ is still only
about $0.2$, which isn't enough for the linear eigenmodes to dominate the turbulent structure.
Also not surprising is that the fastest growing eigenfunctions make up a significant fraction of the energy in the $n=0$ suppressed simulation. Where the linear drift wave instability 
(and the turbulent growth rate) is the strongest (at $m \sim 50$), $R_m \sim 0.5$. The linear physics controls the $n=0$ suppressed simulation, and the linear eigenmode structure certainly asserts itself
in the turbulence, but only to a certain degree ($50\%$). Overall, Fig.~\ref{ratios} shows the relative strength of the linear instabilities compared to nonlinear effects for each of the simulations.
While the nonlinear instability is dominant for all of the simulations other than the $n=0$ suppressed one, the linear instabilities do still act to varying degree in all of the simulations.

\section{Conclusion}
\label{conclusion}

The observation of filamentary $k_\para = 0$ structures is common in many different kinds of experiments and simulations~\cite{rogers2010,kamataki2007,brochard2005}. 
Not surprisingly, the presence of these structures is usually attributed to linear flute-like instabilities such as flow-driven Kelvin-Helmholtz or interchange instabilities rather than to 
nonlinear instabilities~\cite{rogers2010,kamataki2007,brochard2005}.
Due to natural limitations in plasma turbulence experiments, one usually has to resort to indirect evidence to 
gain insight into physical mechanisms of observed turbulence. 
On the other hand, numerical simulations have all spatial and temporal information available which allows one to perform detailed turbulence analyses, such as the energetics analysis
performed in this paper. 
This can make turbulence simulations (provided they are validated on observable data) an important tool for uncovering underlying physical mechanisms.

Through simulation of a particular LAPD experiment, this paper has shown a nonlinear instability to be the driving force behind the turbulence. More acurately, this paper has extended
earlier work~\cite{friedman2012b} that showed this. However, this extension is important because it deals with the matter of axial boundary conditions, and the nonlinear instability
depends on axial wave dynamics, so the boundary conditions could have greatly affected this. And although the various boundary conditions used here do
have significant qualitative and/or quantitative effects on the linear instabilities of the system, they do not affect the turbulence in a significant way.
Specifically, the nonlinear instability that preferentially drives $k_\para = 0$ structures remains robust to the change in boundary conditions. Quantitatively, the sheath, zero-value,
and zero-derivative boundary conditions
do cause the linear instability to be more competitive with the nonlinear instability, but this effect is not large enough to change the qualitative picture.


\begin{acknowledgments}
This research was performed under appointment to the Fusion Energy Sciences Fellowship Program administered by Oak Ridge Institute for
Science and Education under a contract between the U.S. Department of Energy and the Oak Ridge Associated Universities. We would like to thank Prof. Paul Terry for many useful discussions.
\end{acknowledgments}

\appendix

\section{Non-periodic Fourier Decomposition}

\begin{figure}[!htbp]
\includegraphics[]{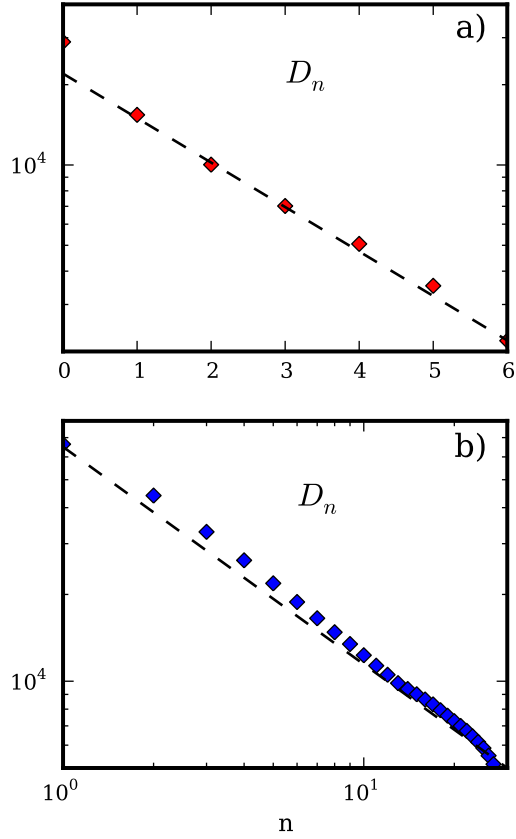}
\hfil
\caption{$D_n$ for \textbf{a)} The simulation with periodic axial
boundaries displaying exponential convergence and \textbf{b)} the simulation with sheath axial boundaries displaying algebraic convergence.}
\label{fourier_convergence}
\end{figure}

It is well known that Fourier reconstructions of signals that contain discontinuities or non-periodic boundaries are subject to Gibbs phenomena. A clear indication of this is the
convergence properties of the Fourier reconstructions. Take a discrete signal with the following Fourier decomposition:

\beq
\label{f_decomp}
f(x) = \sum_{k=-N}^{N} \hat{f}_k e^{2 \pi i k x},
\eeq

where the $\hat{f}_k$ are ordered in the sum by the size of their absolute value with $\hat{f}_0$ being the largest Fourier coefficient. The Fourier reconstruction of order $n<N$ is then:

\beq
\label{f_recon}
g_n(x) = \sum_{k=-n}^{n} \hat{f}_k e^{2 \pi i k x}
\eeq

There are several types of convergences of the $g_n$, one of which is the $L1$ norm. Defining the difference between the original signal and the Fourier reconstruction of order $n$ as
$D_n = \sum_x |f(x) - g_n(x)|$, one can look at the convergence of $D_n$ as a function of $n$. For periodic signals, $D_n$ converges exponentially, while it only converges algebraically
(power law) for non-periodic or discontinuous signals. 

$D_n$ is plotted for the cases of the periodic and sheath simulations in Fig.~\ref{fourier_convergence}. Even though the x-axis label $n$ indicates the mode with the $n^{th}$ largest amplitude
by construction of Eq.~\ref{f_recon}, it also happens to correspond to the axial mode number for all but the last few $n$. In other words, in both simulations, most of the energy is contained
in $n=0$ modes followed by $n=1$ modes and so on. Therefore, in reality, the axial Fourier decomposition is a useful tool for even the sheath simulation despite the fact that Fourier
modes are not natural eigenmodes in this case.


%


\begin{thebibliography}{32}%
\makeatletter
\providecommand \@ifxundefined [1]{%
 \@ifx{#1\undefined}
}%
\providecommand \@ifnum [1]{%
 \ifnum #1\expandafter \@firstoftwo
 \else \expandafter \@secondoftwo
 \fi
}%
\providecommand \@ifx [1]{%
 \ifx #1\expandafter \@firstoftwo
 \else \expandafter \@secondoftwo
 \fi
}%
\providecommand \natexlab [1]{#1}%
\providecommand \enquote  [1]{``#1''}%
\providecommand \bibnamefont  [1]{#1}%
\providecommand \bibfnamefont [1]{#1}%
\providecommand \citenamefont [1]{#1}%
\providecommand \href@noop [0]{\@secondoftwo}%
\providecommand \href [0]{\begingroup \@sanitize@url \@href}%
\providecommand \@href[1]{\@@startlink{#1}\@@href}%
\providecommand \@@href[1]{\endgroup#1\@@endlink}%
\providecommand \@sanitize@url [0]{\catcode `\\12\catcode `\$12\catcode
  `\&12\catcode `\#12\catcode `\^12\catcode `\_12\catcode `\%12\relax}%
\providecommand \@@startlink[1]{}%
\providecommand \@@endlink[0]{}%
\providecommand \url  [0]{\begingroup\@sanitize@url \@url }%
\providecommand \@url [1]{\endgroup\@href {#1}{\urlprefix }}%
\providecommand \urlprefix  [0]{URL }%
\providecommand \Eprint [0]{\href }%
\providecommand \doibase [0]{http://dx.doi.org/}%
\providecommand \selectlanguage [0]{\@gobble}%
\providecommand \bibinfo  [0]{\@secondoftwo}%
\providecommand \bibfield  [0]{\@secondoftwo}%
\providecommand \translation [1]{[#1]}%
\providecommand \BibitemOpen [0]{}%
\providecommand \bibitemStop [0]{}%
\providecommand \bibitemNoStop [0]{.\EOS\space}%
\providecommand \EOS [0]{\spacefactor3000\relax}%
\providecommand \BibitemShut  [1]{\csname bibitem#1\endcsname}%
\let\auto@bib@innerbib\@empty
\bibitem [{\citenamefont {Manneville}(2008)}]{manneville2008}%
  \BibitemOpen
  \bibfield  {author} {\bibinfo {author} {\bibfnamefont {P.}~\bibnamefont
  {Manneville}},\ }\bibfield  {title} {\enquote {\bibinfo {title}
  {Understanding the sub-critical transition to turbulence in wall flows},}\
  }\href@noop {} {\bibfield  {journal} {\bibinfo  {journal} {Pramana}\ }\textbf
  {\bibinfo {volume} {70}},\ \bibinfo {pages} {1009--1021} (\bibinfo {year}
  {2008})}\BibitemShut {NoStop}%
\bibitem [{\citenamefont {Waltz}(1985)}]{waltz1985}%
  \BibitemOpen
  \bibfield  {author} {\bibinfo {author} {\bibfnamefont {R.~E.}\ \bibnamefont
  {Waltz}},\ }\bibfield  {title} {\enquote {\bibinfo {title} {Subcritical
  magnetohydrodynamic turbulence},}\ }\href@noop {} {\bibfield  {journal}
  {\bibinfo  {journal} {Phys. Rev. Lett.}\ }\textbf {\bibinfo {volume} {55}},\
  \bibinfo {pages} {1098} (\bibinfo {year} {1985})}\BibitemShut {NoStop}%
\bibitem [{\citenamefont {Scott}(1990)}]{scott1990}%
  \BibitemOpen
  \bibfield  {author} {\bibinfo {author} {\bibfnamefont {B.~D.}\ \bibnamefont
  {Scott}},\ }\bibfield  {title} {\enquote {\bibinfo {title} {Self-sustained
  collisional drift-wave turbulence in a sheared magnetic field},}\ }\href@noop
  {} {\bibfield  {journal} {\bibinfo  {journal} {Phys. Rev. Lett.}\ }\textbf
  {\bibinfo {volume} {65}},\ \bibinfo {pages} {3289} (\bibinfo {year}
  {1990})}\BibitemShut {NoStop}%
\bibitem [{\citenamefont {Scott}(1992)}]{scott1992}%
  \BibitemOpen
  \bibfield  {author} {\bibinfo {author} {\bibfnamefont {B.~D.}\ \bibnamefont
  {Scott}},\ }\bibfield  {title} {\enquote {\bibinfo {title} {The mechanism of
  self sustainment in collisional drift wave turbulence},}\ }\href@noop {}
  {\bibfield  {journal} {\bibinfo  {journal} {Phys. Fluids B}\ }\textbf
  {\bibinfo {volume} {4}},\ \bibinfo {pages} {2468} (\bibinfo {year}
  {1992})}\BibitemShut {NoStop}%
\bibitem [{\citenamefont {Nordman}, \citenamefont {Pavlenko},\ and\
  \citenamefont {Weiland}(1993)}]{nordman1993}%
  \BibitemOpen
  \bibfield  {author} {\bibinfo {author} {\bibfnamefont {H.}~\bibnamefont
  {Nordman}}, \bibinfo {author} {\bibfnamefont {V.~P.}\ \bibnamefont
  {Pavlenko}}, \ and\ \bibinfo {author} {\bibfnamefont {J.}~\bibnamefont
  {Weiland}},\ }\bibfield  {title} {\enquote {\bibinfo {title} {Subcritical
  reactive drift wave turbulence},}\ }\href@noop {} {\bibfield  {journal}
  {\bibinfo  {journal} {Phys. Fluids B}\ }\textbf {\bibinfo {volume} {5}},\
  \bibinfo {pages} {402} (\bibinfo {year} {1993})}\BibitemShut {NoStop}%
\bibitem [{\citenamefont {Itoh}\ \emph {et~al.}(1996)\citenamefont {Itoh},
  \citenamefont {Itoh}, \citenamefont {Yagi},\ and\ \citenamefont
  {Fukuyama}}]{itoh1996}%
  \BibitemOpen
  \bibfield  {author} {\bibinfo {author} {\bibfnamefont {K.}~\bibnamefont
  {Itoh}}, \bibinfo {author} {\bibfnamefont {S.~I.}\ \bibnamefont {Itoh}},
  \bibinfo {author} {\bibfnamefont {M.}~\bibnamefont {Yagi}}, \ and\ \bibinfo
  {author} {\bibfnamefont {A.}~\bibnamefont {Fukuyama}},\ }\bibfield  {title}
  {\enquote {\bibinfo {title} {Subcritical excitation of plasma turbulence},}\
  }\href@noop {} {\bibfield  {journal} {\bibinfo  {journal} {Journal of the
  Physical Society of Japan}\ }\textbf {\bibinfo {volume} {65}},\ \bibinfo
  {pages} {2749--2752} (\bibinfo {year} {1996})}\BibitemShut {NoStop}%
\bibitem [{\citenamefont {Highcock}\ \emph {et~al.}(2012)\citenamefont
  {Highcock}, \citenamefont {Schekochihin}, \citenamefont {Cowley},
  \citenamefont {Barnes}, \citenamefont {Parra}, \citenamefont {Roach},\ and\
  \citenamefont {Dorland}}]{highcock2012}%
  \BibitemOpen
  \bibfield  {author} {\bibinfo {author} {\bibfnamefont {E.~G.}\ \bibnamefont
  {Highcock}}, \bibinfo {author} {\bibfnamefont {A.~A.}\ \bibnamefont
  {Schekochihin}}, \bibinfo {author} {\bibfnamefont {S.~C.}\ \bibnamefont
  {Cowley}}, \bibinfo {author} {\bibfnamefont {M.}~\bibnamefont {Barnes}},
  \bibinfo {author} {\bibfnamefont {F.~I.}\ \bibnamefont {Parra}}, \bibinfo
  {author} {\bibfnamefont {C.~M.}\ \bibnamefont {Roach}}, \ and\ \bibinfo
  {author} {\bibfnamefont {W.}~\bibnamefont {Dorland}},\ }\bibfield  {title}
  {\enquote {\bibinfo {title} {Zero-turbulence manifold in a toroidal
  plasma},}\ }\href@noop {} {\bibfield  {journal} {\bibinfo  {journal} {Phys.
  Rev. Lett.}\ }\textbf {\bibinfo {volume} {109}},\ \bibinfo {pages} {265001}
  (\bibinfo {year} {2012})}\BibitemShut {NoStop}%
\bibitem [{\citenamefont {Dimits}\ \emph {et~al.}(2000)\citenamefont {Dimits},
  \citenamefont {Bateman}, \citenamefont {Beer}, \citenamefont {Cohen},
  \citenamefont {Dorland}, \citenamefont {Hammett}, \citenamefont {Kim},
  \citenamefont {Kinsey}, \citenamefont {Kotschenreuther}, \citenamefont
  {Kritz}, \citenamefont {Lao}, \citenamefont {Mandrekas}, \citenamefont
  {Nevins}, \citenamefont {Parker}, \citenamefont {Redd}, \citenamefont
  {Shumaker}, \citenamefont {Sydora},\ and\ \citenamefont
  {Weiland}}]{dimits2000}%
  \BibitemOpen
  \bibfield  {author} {\bibinfo {author} {\bibfnamefont {A.~M.}\ \bibnamefont
  {Dimits}}, \bibinfo {author} {\bibfnamefont {G.}~\bibnamefont {Bateman}},
  \bibinfo {author} {\bibfnamefont {M.~A.}\ \bibnamefont {Beer}}, \bibinfo
  {author} {\bibfnamefont {B.~I.}\ \bibnamefont {Cohen}}, \bibinfo {author}
  {\bibfnamefont {W.}~\bibnamefont {Dorland}}, \bibinfo {author} {\bibfnamefont
  {G.~W.}\ \bibnamefont {Hammett}}, \bibinfo {author} {\bibfnamefont
  {C.}~\bibnamefont {Kim}}, \bibinfo {author} {\bibfnamefont {J.~E.}\
  \bibnamefont {Kinsey}}, \bibinfo {author} {\bibfnamefont {M.}~\bibnamefont
  {Kotschenreuther}}, \bibinfo {author} {\bibfnamefont {A.~H.}\ \bibnamefont
  {Kritz}}, \bibinfo {author} {\bibfnamefont {L.~L.}\ \bibnamefont {Lao}},
  \bibinfo {author} {\bibfnamefont {J.}~\bibnamefont {Mandrekas}}, \bibinfo
  {author} {\bibfnamefont {W.~M.}\ \bibnamefont {Nevins}}, \bibinfo {author}
  {\bibfnamefont {S.~E.}\ \bibnamefont {Parker}}, \bibinfo {author}
  {\bibfnamefont {A.~J.}\ \bibnamefont {Redd}}, \bibinfo {author}
  {\bibfnamefont {D.~E.}\ \bibnamefont {Shumaker}}, \bibinfo {author}
  {\bibfnamefont {R.}~\bibnamefont {Sydora}}, \ and\ \bibinfo {author}
  {\bibfnamefont {J.}~\bibnamefont {Weiland}},\ }\bibfield  {title} {\enquote
  {\bibinfo {title} {Comparisons and physics basis of tokamak transport models
  and turbulece simulations},}\ }\href@noop {} {\bibfield  {journal} {\bibinfo
  {journal} {Phys. Plasmas}\ }\textbf {\bibinfo {volume} {7}},\ \bibinfo
  {pages} {969} (\bibinfo {year} {2000})}\BibitemShut {NoStop}%
\bibitem [{\citenamefont {Ernst}\ \emph {et~al.}(2004)\citenamefont {Ernst},
  \citenamefont {Bonoli}, \citenamefont {Catto}, \citenamefont {Dorland},
  \citenamefont {Fiore}, \citenamefont {Granetz}, \citenamefont {Greenwald},
  \citenamefont {Hubbard}, \citenamefont {Prokolab}, \citenamefont {Redi},
  \citenamefont {Rice},\ and\ \citenamefont {Zhurovich}}]{ernst2004}%
  \BibitemOpen
  \bibfield  {author} {\bibinfo {author} {\bibfnamefont {D.~R.}\ \bibnamefont
  {Ernst}}, \bibinfo {author} {\bibfnamefont {P.~T.}\ \bibnamefont {Bonoli}},
  \bibinfo {author} {\bibfnamefont {P.~J.}\ \bibnamefont {Catto}}, \bibinfo
  {author} {\bibfnamefont {W.}~\bibnamefont {Dorland}}, \bibinfo {author}
  {\bibfnamefont {C.~L.}\ \bibnamefont {Fiore}}, \bibinfo {author}
  {\bibfnamefont {R.~S.}\ \bibnamefont {Granetz}}, \bibinfo {author}
  {\bibfnamefont {M.}~\bibnamefont {Greenwald}}, \bibinfo {author}
  {\bibfnamefont {A.~E.}\ \bibnamefont {Hubbard}}, \bibinfo {author}
  {\bibfnamefont {M.}~\bibnamefont {Prokolab}}, \bibinfo {author}
  {\bibfnamefont {M.~H.}\ \bibnamefont {Redi}}, \bibinfo {author}
  {\bibfnamefont {J.~E.}\ \bibnamefont {Rice}}, \ and\ \bibinfo {author}
  {\bibfnamefont {K.}~\bibnamefont {Zhurovich}},\ }\bibfield  {title} {\enquote
  {\bibinfo {title} {Role of trapped electron mode turbulence in internal
  transport barrier control in the alcator c-mod tokamak},}\ }\href@noop {}
  {\bibfield  {journal} {\bibinfo  {journal} {Phys. Plasmas}\ }\textbf
  {\bibinfo {volume} {11}},\ \bibinfo {pages} {2637} (\bibinfo {year}
  {2004})}\BibitemShut {NoStop}%
\bibitem [{\citenamefont {Biskamp}\ and\ \citenamefont
  {Zeiler}(1995)}]{biskamp1995}%
  \BibitemOpen
  \bibfield  {author} {\bibinfo {author} {\bibfnamefont {D.}~\bibnamefont
  {Biskamp}}\ and\ \bibinfo {author} {\bibfnamefont {A.}~\bibnamefont
  {Zeiler}},\ }\bibfield  {title} {\enquote {\bibinfo {title} {Nonlinear
  instability mechanism in 3d collisional drift-wave turbulence},}\ }\href@noop
  {} {\bibfield  {journal} {\bibinfo  {journal} {Phys. Rev. Lett.}\ }\textbf
  {\bibinfo {volume} {74}},\ \bibinfo {pages} {706} (\bibinfo {year}
  {1995})}\BibitemShut {NoStop}%
\bibitem [{\citenamefont {Drake}, \citenamefont {Zeiler},\ and\ \citenamefont
  {Biskamp}(1995)}]{drake1995}%
  \BibitemOpen
  \bibfield  {author} {\bibinfo {author} {\bibfnamefont {J.~F.}\ \bibnamefont
  {Drake}}, \bibinfo {author} {\bibfnamefont {A.}~\bibnamefont {Zeiler}}, \
  and\ \bibinfo {author} {\bibfnamefont {D.}~\bibnamefont {Biskamp}},\
  }\bibfield  {title} {\enquote {\bibinfo {title} {Nonlinear self-sustained
  drift-wave turbulence},}\ }\href@noop {} {\bibfield  {journal} {\bibinfo
  {journal} {Phys. Rev. Lett.}\ }\textbf {\bibinfo {volume} {75}},\ \bibinfo
  {pages} {4222} (\bibinfo {year} {1995})}\BibitemShut {NoStop}%
\bibitem [{\citenamefont {Zeiler}\ \emph {et~al.}(1996)\citenamefont {Zeiler},
  \citenamefont {Biskamp}, \citenamefont {Drake},\ and\ \citenamefont
  {Guzdar}}]{zeiler1996}%
  \BibitemOpen
  \bibfield  {author} {\bibinfo {author} {\bibfnamefont {A.}~\bibnamefont
  {Zeiler}}, \bibinfo {author} {\bibfnamefont {D.}~\bibnamefont {Biskamp}},
  \bibinfo {author} {\bibfnamefont {J.~F.}\ \bibnamefont {Drake}}, \ and\
  \bibinfo {author} {\bibfnamefont {P.~N.}\ \bibnamefont {Guzdar}},\ }\bibfield
   {title} {\enquote {\bibinfo {title} {Three-dimensional fluid simulations of
  tokamak edge turbulence},}\ }\href@noop {} {\bibfield  {journal} {\bibinfo
  {journal} {Phys. Plasmas}\ }\textbf {\bibinfo {volume} {3}},\ \bibinfo
  {pages} {2951} (\bibinfo {year} {1996})}\BibitemShut {NoStop}%
\bibitem [{\citenamefont {Zeiler}, \citenamefont {Drake},\ and\ \citenamefont
  {Biskamp}(1997)}]{zeiler1997}%
  \BibitemOpen
  \bibfield  {author} {\bibinfo {author} {\bibfnamefont {A.}~\bibnamefont
  {Zeiler}}, \bibinfo {author} {\bibfnamefont {J.~F.}\ \bibnamefont {Drake}}, \
  and\ \bibinfo {author} {\bibfnamefont {D.}~\bibnamefont {Biskamp}},\
  }\bibfield  {title} {\enquote {\bibinfo {title} {Electron temperature
  fluctuations in drift-resistive ballooning turbulence},}\ }\href@noop {}
  {\bibfield  {journal} {\bibinfo  {journal} {Phys. Plasmas}\ }\textbf
  {\bibinfo {volume} {4}},\ \bibinfo {pages} {991} (\bibinfo {year}
  {1997})}\BibitemShut {NoStop}%
\bibitem [{\citenamefont {Korsholm}, \citenamefont {Michelsen},\ and\
  \citenamefont {Naulin}(1999)}]{korsholm1999}%
  \BibitemOpen
  \bibfield  {author} {\bibinfo {author} {\bibfnamefont {S.~B.}\ \bibnamefont
  {Korsholm}}, \bibinfo {author} {\bibfnamefont {P.~K.}\ \bibnamefont
  {Michelsen}}, \ and\ \bibinfo {author} {\bibfnamefont {V.}~\bibnamefont
  {Naulin}},\ }\bibfield  {title} {\enquote {\bibinfo {title} {Resistive drift
  wave turbulence in a three-dimensional geometry},}\ }\href@noop {} {\bibfield
   {journal} {\bibinfo  {journal} {Phys. Plasmas}\ }\textbf {\bibinfo {volume}
  {6}},\ \bibinfo {pages} {2401} (\bibinfo {year} {1999})}\BibitemShut
  {NoStop}%
\bibitem [{\citenamefont {Scott}(2002)}]{scott2002}%
  \BibitemOpen
  \bibfield  {author} {\bibinfo {author} {\bibfnamefont {B.~D.}\ \bibnamefont
  {Scott}},\ }\bibfield  {title} {\enquote {\bibinfo {title} {The nonlinear
  drift wave instability and its role in tokamak edge turbulence},}\
  }\href@noop {} {\bibfield  {journal} {\bibinfo  {journal} {New J. Physics}\
  }\textbf {\bibinfo {volume} {4}},\ \bibinfo {pages} {52.1--52.30} (\bibinfo
  {year} {2002})}\BibitemShut {NoStop}%
\bibitem [{\citenamefont {Scott}(2003)}]{scott2003}%
  \BibitemOpen
  \bibfield  {author} {\bibinfo {author} {\bibfnamefont {B.~D.}\ \bibnamefont
  {Scott}},\ }\bibfield  {title} {\enquote {\bibinfo {title} {Computation of
  electromagnetic turbulence and anomalous transport mechanisms in tokamak
  plasmas},}\ }\href@noop {} {\bibfield  {journal} {\bibinfo  {journal} {Plasma
  Phys. Control. Fusion}\ }\textbf {\bibinfo {volume} {45}},\ \bibinfo {pages}
  {A385--A398} (\bibinfo {year} {2003})}\BibitemShut {NoStop}%
\bibitem [{\citenamefont {Scott}(2005)}]{scott2005}%
  \BibitemOpen
  \bibfield  {author} {\bibinfo {author} {\bibfnamefont {B.~D.}\ \bibnamefont
  {Scott}},\ }\bibfield  {title} {\enquote {\bibinfo {title} {Drift wave versus
  interchange turbulence in tokamak geometry: Linear versus nonlinear mode
  structure},}\ }\href@noop {} {\bibfield  {journal} {\bibinfo  {journal}
  {Phys. Plasmas}\ }\textbf {\bibinfo {volume} {12}},\ \bibinfo {pages}
  {062314} (\bibinfo {year} {2005})}\BibitemShut {NoStop}%
\bibitem [{\citenamefont {Friedman}\ \emph {et~al.}(2012)\citenamefont
  {Friedman}, \citenamefont {Carter}, \citenamefont {Umansky}, \citenamefont
  {Schaffner},\ and\ \citenamefont {Dudson}}]{friedman2012b}%
  \BibitemOpen
  \bibfield  {author} {\bibinfo {author} {\bibfnamefont {B.}~\bibnamefont
  {Friedman}}, \bibinfo {author} {\bibfnamefont {T.~A.}\ \bibnamefont
  {Carter}}, \bibinfo {author} {\bibfnamefont {M.~V.}\ \bibnamefont {Umansky}},
  \bibinfo {author} {\bibfnamefont {D.}~\bibnamefont {Schaffner}}, \ and\
  \bibinfo {author} {\bibfnamefont {B.}~\bibnamefont {Dudson}},\ }\bibfield
  {title} {\enquote {\bibinfo {title} {Energy dynamics in a simulation of lapd
  turbulence},}\ }\href@noop {} {\bibfield  {journal} {\bibinfo  {journal}
  {Phys. Plasmas}\ }\textbf {\bibinfo {volume} {19}},\ \bibinfo {pages}
  {102307} (\bibinfo {year} {2012})}\BibitemShut {NoStop}%
\bibitem [{\citenamefont {Gekelman}\ \emph {et~al.}(1991)\citenamefont
  {Gekelman}, \citenamefont {Pfister}, \citenamefont {Lucky}, \citenamefont
  {Bamber}, \citenamefont {Leneman},\ and\ \citenamefont
  {Maggs}}]{Gekelman1991}%
  \BibitemOpen
  \bibfield  {author} {\bibinfo {author} {\bibfnamefont {W.}~\bibnamefont
  {Gekelman}}, \bibinfo {author} {\bibfnamefont {H.}~\bibnamefont {Pfister}},
  \bibinfo {author} {\bibfnamefont {Z.}~\bibnamefont {Lucky}}, \bibinfo
  {author} {\bibfnamefont {J.}~\bibnamefont {Bamber}}, \bibinfo {author}
  {\bibfnamefont {D.}~\bibnamefont {Leneman}}, \ and\ \bibinfo {author}
  {\bibfnamefont {J.}~\bibnamefont {Maggs}},\ }\bibfield  {title} {\enquote
  {\bibinfo {title} {Design, construction and properties of the large plasma
  research device - the lapd at ucla},}\ }\href@noop {} {\bibfield  {journal}
  {\bibinfo  {journal} {Rev. Sci. Inst.}\ }\textbf {\bibinfo {volume} {62}},\
  \bibinfo {pages} {2875} (\bibinfo {year} {1991})}\BibitemShut {NoStop}%
\bibitem [{\citenamefont {Berk}, \citenamefont {Ryutov},\ and\ \citenamefont
  {Tsidulko}(1991)}]{berk1991}%
  \BibitemOpen
  \bibfield  {author} {\bibinfo {author} {\bibfnamefont {H.~L.}\ \bibnamefont
  {Berk}}, \bibinfo {author} {\bibfnamefont {D.~D.}\ \bibnamefont {Ryutov}}, \
  and\ \bibinfo {author} {\bibfnamefont {Y.~A.}\ \bibnamefont {Tsidulko}},\
  }\bibfield  {title} {\enquote {\bibinfo {title} {Temperature-gradient
  instability induced by conducting end walls},}\ }\href@noop {} {\bibfield
  {journal} {\bibinfo  {journal} {Phys. Fluids B}\ }\textbf {\bibinfo {volume}
  {3}},\ \bibinfo {pages} {1346} (\bibinfo {year} {1991})}\BibitemShut
  {NoStop}%
\bibitem [{\citenamefont {Schaffner}\ \emph {et~al.}(2012)\citenamefont
  {Schaffner}, \citenamefont {Carter}, \citenamefont {Rossi}, \citenamefont
  {Guice}, \citenamefont {Maggs}, \citenamefont {Vincena},\ and\ \citenamefont
  {Friedman}}]{schaffner2012}%
  \BibitemOpen
  \bibfield  {author} {\bibinfo {author} {\bibfnamefont {D.~A.}\ \bibnamefont
  {Schaffner}}, \bibinfo {author} {\bibfnamefont {T.~A.}\ \bibnamefont
  {Carter}}, \bibinfo {author} {\bibfnamefont {G.~D.}\ \bibnamefont {Rossi}},
  \bibinfo {author} {\bibfnamefont {D.~S.}\ \bibnamefont {Guice}}, \bibinfo
  {author} {\bibfnamefont {J.~E.}\ \bibnamefont {Maggs}}, \bibinfo {author}
  {\bibfnamefont {S.}~\bibnamefont {Vincena}}, \ and\ \bibinfo {author}
  {\bibfnamefont {B.}~\bibnamefont {Friedman}},\ }\bibfield  {title} {\enquote
  {\bibinfo {title} {Modification of turbulent transport with continuous
  variation of flow shear in the large plasma device},}\ }\href@noop {}
  {\bibfield  {journal} {\bibinfo  {journal} {Phys. Rev. Lett.}\ }\textbf
  {\bibinfo {volume} {109}},\ \bibinfo {pages} {135002} (\bibinfo {year}
  {2012})}\BibitemShut {NoStop}%
\bibitem [{\citenamefont {Braginskii}(1965)}]{Braginskii1965}%
  \BibitemOpen
  \bibfield  {author} {\bibinfo {author} {\bibfnamefont {S.~I.}\ \bibnamefont
  {Braginskii}},\ }\bibfield  {title} {\enquote {\bibinfo {title} {T{ransport
  processes in a plasma}},}\ }in\ \href@noop {} {\emph {\bibinfo {booktitle}
  {Reviews of Plasma Physics}}},\ Vol.~\bibinfo {volume} {1},\ \bibinfo
  {editor} {edited by\ \bibinfo {editor} {\bibfnamefont {M.~A.}\ \bibnamefont
  {Leontovich}}}\ (\bibinfo  {publisher} {Consultants Bureau, New York},\
  \bibinfo {year} {1965})\ pp.\ \bibinfo {pages} {205--311}\BibitemShut
  {NoStop}%
\bibitem [{\citenamefont {Dudson}\ \emph {et~al.}(2009)\citenamefont {Dudson},
  \citenamefont {Umansky}, \citenamefont {Xu}, \citenamefont {Snyder},\ and\
  \citenamefont {Wilson}}]{dudson2009}%
  \BibitemOpen
  \bibfield  {author} {\bibinfo {author} {\bibfnamefont {B.~D.}\ \bibnamefont
  {Dudson}}, \bibinfo {author} {\bibfnamefont {M.~V.}\ \bibnamefont {Umansky}},
  \bibinfo {author} {\bibfnamefont {X.~Q.}\ \bibnamefont {Xu}}, \bibinfo
  {author} {\bibfnamefont {P.~B.}\ \bibnamefont {Snyder}}, \ and\ \bibinfo
  {author} {\bibfnamefont {H.~R.}\ \bibnamefont {Wilson}},\ }\bibfield  {title}
  {\enquote {\bibinfo {title} {Bout++: A framework for parallel plasma fluid
  simulations.}}\ }\href@noop {} {\bibfield  {journal} {\bibinfo  {journal}
  {Computer Physics Communications}\ ,\ \bibinfo {pages} {1467--1480}}
  (\bibinfo {year} {2009})}\BibitemShut {NoStop}%
\bibitem [{\citenamefont {Popovich}\ \emph
  {et~al.}(2010{\natexlab{a}})\citenamefont {Popovich}, \citenamefont
  {Umansky}, \citenamefont {Carter},\ and\ \citenamefont
  {Friedman}}]{Popovich2010a}%
  \BibitemOpen
  \bibfield  {author} {\bibinfo {author} {\bibfnamefont {P.}~\bibnamefont
  {Popovich}}, \bibinfo {author} {\bibfnamefont {M.~V.}\ \bibnamefont
  {Umansky}}, \bibinfo {author} {\bibfnamefont {T.~A.}\ \bibnamefont {Carter}},
  \ and\ \bibinfo {author} {\bibfnamefont {B.}~\bibnamefont {Friedman}},\
  }\bibfield  {title} {\enquote {\bibinfo {title} {Analysis of plasma
  instabilities and verification of bout code for linear plasma device},}\
  }\href@noop {} {\bibfield  {journal} {\bibinfo  {journal} {Phys. Plasmas}\
  }\textbf {\bibinfo {volume} {17}},\ \bibinfo {pages} {102107} (\bibinfo
  {year} {2010}{\natexlab{a}})}\BibitemShut {NoStop}%
\bibitem [{\citenamefont {Popovich}\ \emph
  {et~al.}(2010{\natexlab{b}})\citenamefont {Popovich}, \citenamefont
  {Umansky}, \citenamefont {Carter},\ and\ \citenamefont
  {Friedman}}]{Popovich2010b}%
  \BibitemOpen
  \bibfield  {author} {\bibinfo {author} {\bibfnamefont {P.}~\bibnamefont
  {Popovich}}, \bibinfo {author} {\bibfnamefont {M.~V.}\ \bibnamefont
  {Umansky}}, \bibinfo {author} {\bibfnamefont {T.~A.}\ \bibnamefont {Carter}},
  \ and\ \bibinfo {author} {\bibfnamefont {B.}~\bibnamefont {Friedman}},\
  }\bibfield  {title} {\enquote {\bibinfo {title} {Modeling of plasma
  turbulence and transport in the large plasma device},}\ }\href@noop {}
  {\bibfield  {journal} {\bibinfo  {journal} {Phys. Plasmas}\ }\textbf
  {\bibinfo {volume} {17}},\ \bibinfo {pages} {122312} (\bibinfo {year}
  {2010}{\natexlab{b}})}\BibitemShut {NoStop}%
\bibitem [{\citenamefont {Umansky}\ \emph {et~al.}(2011)\citenamefont
  {Umansky}, \citenamefont {Popovich}, \citenamefont {Carter}, \citenamefont
  {Friedman},\ and\ \citenamefont {Nevins}}]{Umansky2011}%
  \BibitemOpen
  \bibfield  {author} {\bibinfo {author} {\bibfnamefont {M.~V.}\ \bibnamefont
  {Umansky}}, \bibinfo {author} {\bibfnamefont {P.}~\bibnamefont {Popovich}},
  \bibinfo {author} {\bibfnamefont {T.~A.}\ \bibnamefont {Carter}}, \bibinfo
  {author} {\bibfnamefont {B.}~\bibnamefont {Friedman}}, \ and\ \bibinfo
  {author} {\bibfnamefont {W.~M.}\ \bibnamefont {Nevins}},\ }\bibfield  {title}
  {\enquote {\bibinfo {title} {Numerical simulation and analysis of plasma
  turbulence the large plasma device},}\ }\href@noop {} {\bibfield  {journal}
  {\bibinfo  {journal} {Phys. Plasmas}\ }\textbf {\bibinfo {volume} {18}},\
  \bibinfo {pages} {055709} (\bibinfo {year} {2011})}\BibitemShut {NoStop}%
\bibitem [{\citenamefont {Friedman}, \citenamefont {Umansky},\ and\
  \citenamefont {Carter}(2012)}]{friedman2012}%
  \BibitemOpen
  \bibfield  {author} {\bibinfo {author} {\bibfnamefont {B.}~\bibnamefont
  {Friedman}}, \bibinfo {author} {\bibfnamefont {M.~V.}\ \bibnamefont
  {Umansky}}, \ and\ \bibinfo {author} {\bibfnamefont {T.~A.}\ \bibnamefont
  {Carter}},\ }\bibfield  {title} {\enquote {\bibinfo {title} {Grid convergence
  study in a simulation of lapd turbulence},}\ }\href@noop {} {\bibfield
  {journal} {\bibinfo  {journal} {Contrib. Plasma Phys.}\ }\textbf {\bibinfo
  {volume} {52}},\ \bibinfo {pages} {412--416} (\bibinfo {year}
  {2012})}\BibitemShut {NoStop}%
\bibitem [{\citenamefont {Loizu}\ \emph {et~al.}(2012)\citenamefont {Loizu},
  \citenamefont {Ricci}, \citenamefont {Halpern},\ and\ \citenamefont
  {Jolliet}}]{loizu2012}%
  \BibitemOpen
  \bibfield  {author} {\bibinfo {author} {\bibfnamefont {J.}~\bibnamefont
  {Loizu}}, \bibinfo {author} {\bibfnamefont {P.}~\bibnamefont {Ricci}},
  \bibinfo {author} {\bibfnamefont {F.}~\bibnamefont {Halpern}}, \ and\
  \bibinfo {author} {\bibfnamefont {S.}~\bibnamefont {Jolliet}},\ }\bibfield
  {title} {\enquote {\bibinfo {title} {Boundary conditions for plasma fluid
  models at the magnetic presheath entrance},}\ }\href@noop {} {\bibfield
  {journal} {\bibinfo  {journal} {Phys. Plasmas}\ }\textbf {\bibinfo {volume}
  {13}},\ \bibinfo {pages} {122307} (\bibinfo {year} {2012})}\BibitemShut
  {NoStop}%
\bibitem [{\citenamefont {Hatch}\ \emph {et~al.}(2011)\citenamefont {Hatch},
  \citenamefont {Terry}, \citenamefont {Jenko}, \citenamefont {Merz},\ and\
  \citenamefont {Nevins}}]{hatch2011}%
  \BibitemOpen
  \bibfield  {author} {\bibinfo {author} {\bibfnamefont {D.~R.}\ \bibnamefont
  {Hatch}}, \bibinfo {author} {\bibfnamefont {P.~W.}\ \bibnamefont {Terry}},
  \bibinfo {author} {\bibfnamefont {F.}~\bibnamefont {Jenko}}, \bibinfo
  {author} {\bibfnamefont {F.}~\bibnamefont {Merz}}, \ and\ \bibinfo {author}
  {\bibfnamefont {W.~M.}\ \bibnamefont {Nevins}},\ }\bibfield  {title}
  {\enquote {\bibinfo {title} {Saturation of gyrokinetic turbulence through
  damped eigenmodes},}\ }\href@noop {} {\bibfield  {journal} {\bibinfo
  {journal} {Phys. Rev. Lett.}\ }\textbf {\bibinfo {volume} {106}},\ \bibinfo
  {pages} {115003} (\bibinfo {year} {2011})}\BibitemShut {NoStop}%
\bibitem [{\citenamefont {Rogers}\ and\ \citenamefont
  {Ricci}(2010)}]{rogers2010}%
  \BibitemOpen
  \bibfield  {author} {\bibinfo {author} {\bibfnamefont {B.}~\bibnamefont
  {Rogers}}\ and\ \bibinfo {author} {\bibfnamefont {P.}~\bibnamefont {Ricci}},\
  }\bibfield  {title} {\enquote {\bibinfo {title} {Low-frequency turbulence in
  a linear magnetized plasma},}\ }\href@noop {} {\bibfield  {journal} {\bibinfo
   {journal} {Phys. Rev. Lett.}\ }\textbf {\bibinfo {volume} {104}},\ \bibinfo
  {pages} {225002} (\bibinfo {year} {2010})}\BibitemShut {NoStop}%
\bibitem [{\citenamefont {Kamataki}\ \emph {et~al.}(2007)\citenamefont
  {Kamataki}, \citenamefont {Nagashima}, \citenamefont {Shinohara},
  \citenamefont {Kawai}, \citenamefont {Yagi}, \citenamefont {Itoh},\ and\
  \citenamefont {Itoh}}]{kamataki2007}%
  \BibitemOpen
  \bibfield  {author} {\bibinfo {author} {\bibfnamefont {K.}~\bibnamefont
  {Kamataki}}, \bibinfo {author} {\bibfnamefont {Y.}~\bibnamefont {Nagashima}},
  \bibinfo {author} {\bibfnamefont {S.}~\bibnamefont {Shinohara}}, \bibinfo
  {author} {\bibfnamefont {Y.}~\bibnamefont {Kawai}}, \bibinfo {author}
  {\bibfnamefont {M.}~\bibnamefont {Yagi}}, \bibinfo {author} {\bibfnamefont
  {K.}~\bibnamefont {Itoh}}, \ and\ \bibinfo {author} {\bibfnamefont {S.~I.}\
  \bibnamefont {Itoh}},\ }\bibfield  {title} {\enquote {\bibinfo {title}
  {Coexistence of collisional drift and flute wave instabilities in bounded
  linear ecr plasma},}\ }\href@noop {} {\bibfield  {journal} {\bibinfo
  {journal} {J. Physical Society of Japan}\ }\textbf {\bibinfo {volume} {76}},\
  \bibinfo {pages} {054501} (\bibinfo {year} {2007})}\BibitemShut {NoStop}%
\bibitem [{\citenamefont {Brochard}, \citenamefont {Gravier},\ and\
  \citenamefont {Bonhomme}(2005)}]{brochard2005}%
  \BibitemOpen
  \bibfield  {author} {\bibinfo {author} {\bibfnamefont {F.}~\bibnamefont
  {Brochard}}, \bibinfo {author} {\bibfnamefont {E.}~\bibnamefont {Gravier}}, \
  and\ \bibinfo {author} {\bibfnamefont {G.}~\bibnamefont {Bonhomme}},\
  }\bibfield  {title} {\enquote {\bibinfo {title} {Transition from flute modes
  to drift waves in a magnetized plasma column},}\ }\href@noop {} {\bibfield
  {journal} {\bibinfo  {journal} {Phys. Plasmas}\ }\textbf {\bibinfo {volume}
  {12}},\ \bibinfo {pages} {062104} (\bibinfo {year} {2005})}\BibitemShut
  {NoStop}%
\end{thebibliography}
\end{document}